# NEW LOSSES MECHANISM IN GRAPHENE NANORESONATORS DUE TO THE SYNTHETIC ELECTRIC FIELDS CAUSED BY INHERENT OUT-OF- PLAIN MEMBRANE CORRUGATIONS

Natalie E. Firsova

Institute for Problems of Mechanical Engineering,
the Russian Academy of Sciences, St. Petersburg 199178, Russia, nef2@mail.ru ,

Yuriy A. Firsov

A.F.Ioffe Physical-Technical Institute,
the Russian Academy of Sciences, St.Petersburg, Russia, yuafirsov@rambler.ru

**New losses mechanism in monolayer graphene nanoresonators caused by dissipative intravalley currents stipulated by the synthetic electric fields is considered. These fields are generated by time-dependent gauge fields arising in graphene membrane due to its intrinsic out-of- plain distortions and the influence of the external periodic electromotive force. This losses mechanism accounts for essential part (about 40 percents) of losses in graphene nanoresonator and is specific just for graphene.**

**The ways of the minimization of this kind of dissipation (increase of the quality factor of the electromechanical system) are discussed. It is explained why one can increase quality factor by correctly chosen combination of strains (by strain engineering). Besides, it is shown that quality factor can be increased by switching on a magnetic field perpendicular to graphene membrane.**

## Introduction

The recent successful preparation of one-atom layer of carbons, graphene [1,2], gave rise to the development of the $2D$-physics. However, the question whether a strictly 2D crystal can exist was first raised theoretically more than 70 years ago by Peierls [3,4] and Landau [6,7]. They showed that in the standard harmonic approximation, thermal fluctuations should destroy long range order, essentially resulting " melting" of a 2D lattice at any finite temperature. Mermin and Wagner proved that any long-range order could not exist in one and two dimensions. The same is true for crystalline order in 2D [5]. Really all the observed mono-atomic graphene samples have inherent stable corrugations , i.e, out-of-plain deformations (ripples, bubbles, wrinkles etc.), see for instance [6], where it was discovered that "….graphene sheets are not perfectly flat but exhibit intrinsic microscopic roughening…" and also "…the observed corrugations in the third dimension may shed light on subtle reasons behind the stability of $2D$-crystals"). In [7] it was theoretically shown the divergency mentioned above caused "dangerous" fluctuations can, however, be suppressed by anharmonic coupling between bending and stretching modes making that a two-dimensional membrane can exist but should present strong height thermal fluctuations (about 7nm) and ripples spontaneously appear. So considering graphene membrane with distortions we do not study a specific case but the general one. These corrugations lead to the arising of pseudomagnetic field (gauge field), see for instance review [8]. These inevitably existing fields in graphene are about several Tesla.

It is interesting that such fields not long ago were artificially created in another nontrivial system. It was done in rubidium Bose-Einstein condensate (BEC). This field was produced as time-dependent which led to the appearance of the so-called synthetic electric fields [9]. In [9] effective time-dependent vector-potential for neutral atoms was created via interaction with laser light, generated synthetic electric field simulating charged condensed matter system with neutral atoms.

Synthetic electric field arises also in graphene when pseudomagnetic gauge field is generated by time-dependent distortions such as flexural phonons [10] or due to the influence of the external time-dependent electromotive force in such devices as graphene nanoresonators, And we consider below just this case.

Nanoresonators proved to be very useful in a great number of applications in different spheres of activity.

In series of new small-size devices named nanoelectromechanical systems (NEMS) (see [11], [12]) the nanoresonators seem to be especially perspective. At first for the fabrication of nanoresonators such materials were used as piezoelectrics, silicon, metallic nanowires, carbon nanotubes. The best dynamic characteristics may be achieved as the resonator size and mass scaled down (which is assumed in classical linear elastic Bernoulli-Euler beam theory). Resonance frequency may be essentially increased while the quality factor $Q$ will not become much worse (for instance see [13,14]). This allows the sensitive detection of many physical properties such as quantum state, spin, force, molecular mass. These possibilities opened new investigations in biology : virus, protein, and DNA detection, detection of enzymatic activity etc

New opportunities arise if we come to such material as graphene – one carbon atom layer. For instance, recently a new especially precise method was suggested for mass detection (with zg sensitivity) based on NEM mass spectrometer [15] exploiting the advantage of graphene membranes.

Different modifications of graphene nanoresonators were studied, for instance in [16-18]. It was shown that the damping rate increases linearly with resonance frequency. Different kinds of loss mechanisms are discussed in [16-21]. Some of them are common to all experimental setups: attachment losses, thermoelastic dissipation etc. The others depend on actuation scheme, for instance, magnetomotive actuation scheme, capacitive coupling etc. surface-relative losses usually can be modeled by distribution of effective two-level systems. All these possibilities were considered in details in [10 ]. Authors of [10 ] pointed out that in graphene nanoresonators dissipation is dominated by electrostatically coupled graphene layer and doped metallic backgate, the energy being dissipated by increasing electron-hole excitations and due to interaction of charge fluctuation with lower-energy flexural phonons.

However as it was mentioned above such approach does not take into account very specific properties of $2D$-systems. The thing is that in graphene significant role play gauge pseudo-magnetic fields [10] created due to spontaneous generation of large-scale stable distortion of $2D$- graphene surface (ripples, wrinkles…) responsible for its high bending rigidity.

Analytical formulae for pseudo-vector potential $\vec{A}$ for monolayer graphene sheet were obtained for the first time by authors of paper [23] (see also [22]). There exist expectations that these pseudo-magnetic fields can be used for the creation of new graphene nanoelectromechanics. Later it was discovered that these gauge fields may be varied by applying of external strains [24-26] (strain engineering).

However, only in [10] it was pointed out that in graphene one should also take into account that so called synthetic electric fields should arise if pseudo-magnetic gauge fields turn to be time-dependent. Having this idea in mind, authors of [10] calculated damping rates for flexural phonons, their dissipation being caused by these electric fields and associated with them currents (Joule heating).

We'll consider synthetic electric fields which inevitably arise during nanoresonator vibrations driven by external electromotive force. We'll also estimate the resonator intrinsic losses (quality factor $Q$) which these fields cause by heating. We'll show that the corresponding contribution in $1/Q$ is very essential and leads to rather large Joule type losses in graphene nanoresonators.

Of course, the role of synthetic electric fields in other NEMS may be also important.

In the last section of our paper we discuss the methods of graphene nanoresonators Joule type losses reduction .

### *The Model*

For monolayer graphene membrane, the equation of surface being $z = h(x, y)$, for any atom the vectors directed to three nearest neighbors have the form (see for instance [23])

$$\vec{u}_1 = a(\sqrt{3}/2, 1/2), \ \vec{u}_2 = a(-\sqrt{3}/2, 1/2), \ \vec{u}_3 = a(0, -1),$$

where $a = 2,5 A$ is a distance between nearest neighbors in the lattice; $h = h(x, y)$ is a distance from a point $(x, y)$ in the plane $XOY$ to the membrane.

In paper [23] the following formulae for gauge field vector potential $\vec{A}$ are obtained

$$A_x(\vec{r}) + iA_y(\vec{r}) = -\sum_j \delta t_j(\vec{r})e^{i\vec{u}_j \vec{K}} = -\frac{\epsilon_1}{2}\sum_j \left[(\vec{u}_j \cdot \nabla)\nabla h\right]^2 e^{i\vec{u}_j \vec{K}} \quad (1)$$

$$A_x = -\frac{1}{2}A^0\left[(h_{xx})^2 - (h_{yy})^2\right]a^2, \quad A_y = A^0\left[h_{xy}(h_{xx} + h_{yy})\right]a^2, \quad (2)$$

$$A^0 = 3/4 \cdot \epsilon_1/e \cdot c/V_F \quad (3)$$

Here $\epsilon_1 = 2,89 ev$, $\vec{K} = a^{-1}(4\pi/3\sqrt{3}, 0)$ is a Dirac point and $t_j$ - exchange integral with the $j$-th nearest neighbor $j = 1,2,3$, and $A^0$ has the same dimensionality as vector potential. Products of the expressions in square brackets in formulae for $A_x, A_y$ in (2) by $a^2$ are dimensionless, i.e. they are numerical coefficients, their magnitudes being dependent on the deflection depth of the graphene membrane (we take into consideration large-scale deformations such as ripples, wrinkles etc.) and also on the lattice constant value for the current moment of time.

When switching of alternating electromotive field along the $OZ$ axes the vectors $\vec{u}_j$ should get a time depending variation $\Delta \vec{u}_j(t)$ which is proportional to $E_0 \sin \omega t$, i.e. in linear approximation

$$a(t) = a_0 + \Delta a(t) \quad (4)$$

where $a_0 = 2,5 A$ is initial value of the parameter "$a$" at $t = 0$ and

$$\Delta a(t) = \eta_1 E_0 \sin \omega t = a_{00} \sin \omega t \quad (5)$$

Here coefficient $\eta_1$ has dimensionality [cm$^2$/v].

Similarly we assume

$$h(x, y, t) = h_0(x, y) + \Delta h(t) \quad (6)$$

$$\Delta h(t) = \eta_2 E_0 \sin \omega t \cdot \cos(\pi x/2L) = h_{00} \sin \omega t \cdot \cos(\pi x/2L), \quad (7)$$

where $z = h_0(x, y)$ is an equation of the initial membrane surface form and $\eta_2$ has the same dimensionality as $\eta_1$. Both of them describe interaction with actuating field on the microscopic level. The coefficients $\eta_1$, $\eta_2$ may generally speaking depend on $x, y$, but it does not influence the main results of our paper.

Last factor in (7) is connected with the clumping of the opposite membrane edges by $x = \pm L$ (doubly clumped).

Notice that as it is shown in [17] for linear approximation to be reasonable the deflection of graphene nanoresonator vibrations should not be more than 1,1nm. As we assume in our calculations below it is equal to 1nm. Therefore our assumption about linearity is quite reasonable. By the way nonlinear problem was investigated as well in a number of works (see [27] and referencies therein) however we study here but linear case.

In presence of the external actuating periodic electric field $E_0 \sin \omega t$ the gauge field vector-potential $\vec{A}$ will depend on time, i.e. in monolayer graphene membrane the so called synthetic electric field will arise

$$\vec{E}_{syn} = -c^{-1}\vec{A}_t \qquad (8)$$

Let $\omega \approx \omega_{res}$ where $\omega_{res}$ is an eigenfrequency of our resonator. Then substituting (2) - (7) into (8) we find

$$\left(\vec{E}_{syn}\right)_x = -c^{-1}\left(\vec{A}_x\right)_t = A^0/c \cdot \{[(h_{xx}^2 - h_{yy}^2)(\Delta a)_t + ah_{xx}(\Delta h)_{xxt}]a\}, \qquad (9)$$

$$\left(\vec{E}_{syn}\right)_y = -c^{-1}\left(\vec{A}_y\right)_t = -A^0/c \cdot \{h_{xy}[2(h_{xx} + h_{yy})(\Delta a)_t + a(\Delta h)_{xxt}]a\} \qquad (10)$$

Using (5), (7) we get

$$\left(\vec{E}_{syn}\right)_x = (E_0\eta_2)E^0(\omega) \cdot$$
$$\{[(\eta_1/\eta_2)(h_{xx}^2 - h_{yy}^2) - ah_{xx}(\pi/2L)^2 \cos(\pi x/2L)]a\} \cdot \cos \omega t \qquad (11)$$

$$\left(\vec{E}_{syn}\right)_y = -(E_0\eta_2)E^0(\omega) \cdot$$
$$\{h_{xy}[2(\eta_1/\eta_2)(h_{xx} + h_{yy}) - a(\pi/2L)^2 \cos(\pi x/2L)]a\} \cdot \cos \omega t \qquad (12)$$

$$E^0(\omega) = 3/4 \cdot \epsilon_1/e \cdot \omega/V_F \qquad (13)$$

We can write formulae (11), (12) in the form

$$\left(\vec{E}_{syn}\right)_x = E^0(\omega)h_{00}I_x \cos \omega t, \quad \left(\vec{E}_{syn}\right)_y = E^0(\omega)h_{00}I_y \cos \omega t \qquad (14)$$

where $h_{00} = (E_0\eta_2)$ is a resonator oscillation amplitude (deflection) and

$$I_x = \{[(\eta_1/\eta_2)(h_{xx}^2 - h_{yy}^2) - ah_{xx}(\pi/2L)^2 \cos(\pi x/2L)]a\} \qquad (15)$$

$$I_y = \{h_{xy}[2(\eta_1/\eta_2)(h_{xx} + h_{yy}) - a(\pi/2L)^2 \cos(\pi x/2L)]a\} \qquad (16)$$

Remark that dimensionless quantities $I_x, I_y$ do not turn to zero even by zero deflection because of the presence in graphene of such deformations as ripples, wrinkles and so on.

It follows from (14) – (16) that after time averaging we have

$$(\vec{E}_{syn})^2 = (\vec{E}_{syn})_x^2 + (\vec{E}_{syn})_y^2 = (E^0(\omega))^2 h_{00}^2 (I_x^2 + I_y^2)/2 \qquad (17)$$

Now using classical formula we can write Joule type losses $\Delta E_J$ for the period $T = 2\pi/\omega$ in the form

$$\Delta E_J \approx 2\pi (\vec{E}_{syn})^2 \sigma L_x L_y/\omega \qquad (18)$$

where $L_x, L_y$ – membrane sizes. Here generally speaking conductivity $\sigma$ depends on intervalley scattering parameters. But we do not analyze here this dependence. We call this dissipation mechanism "Joule type" as it exists due to intrinsic synthetic field which has quantum origin. Note that in [10] Joule like formula in the problem of obtaining damping rate for flexural phonons was proved on the basis of Kubo formula. (see (14) in [10]). Besides it was shown in [10] (see formula (45) and discussion thereabout) that in graphene twodimensional conductivity $\sigma$ does not (or weakly) depend on activating field frequency. But estimating in the next point below the approximate value of Joule type losses we shall take measured value for $\sigma$ using experimental data. So we find

$$\Delta E_J \approx \pi (E^0(\omega))^2 h_{00}^2 (I_x^2 + I_y^2) \sigma L_x L_y/\omega \qquad (19)$$

Remark that in (18), (19) we took into consideration only chargeless ($div\vec{E}_{syn} = div\,\vec{A}_t = 0$) synthetic electric fields which unlike potential fields are not screened by electrons (see. [10]) and therefore their contribution dominates. Besides in (18), (19) only contribution from one Dirac cone (only one valley, i.e. only one sublattice) is taken into consideration. But as graphene lattice consists of two sublattices (two valleys) we should consider also the field from the second valley. In an ideal case i.e. if there is time-reversal symmetry, [28], these fields have opposite directions and equal magnitudes, and the two valley currents compensate each other. However this question was analysed in [10] where it was shown that the two corresponding valley currents do not compensate each other if we take into account intervalley Coulomb drag effect and intervalley scattering on short range impurities.

From formulae (13), (19) we see that the damping rate linearly depends on frequency. It is interesting that in nanoresonators on the basis of carbon nanotubes the dissipation mechanism connected with electron tunneling through a vibrating nanotube also gives damping rate linearly depending on frequency, [21].

General losses include different nature parts,

$$Q^{-1} = Q_0^{-1} + Q_J^{-1} \qquad (20)$$

Here $Q_0^{-1}$ is connected with dissipation mechanisms studied earlier by other authors (see for instance [16- 21]), and $Q_J^{-1}$ was at first considered and analyzed in the present paper.

We introduce quality factor $Q_J$ connected with Joule type losses as follows

$$Q_J^{-1} = \Delta E_J / E_{total} \tag{21}$$

Here $\Delta E_J$ was found in (19), and the total energy is defined as follows

$$E_{total} = N \cdot m_{at} \cdot \omega^2 \cdot h_{00}^2, \quad N = L_x L_y / (a^2 3\sqrt{3}/2),$$

where $N$ is a number of atoms in graphene membrane, $m_{at}$ –is one atom mass and $h_{00}$ - membrane oscillation amplitude. So we obtain

$$Q_J^{-1} = \pi \frac{3\sqrt{3}}{2} \cdot \frac{\left(E^0(\omega)\right)^2 \cdot \sigma \cdot [a^2(I_x^2 + I_y^2)]}{\omega^3 m_{at}} \tag{22}$$

**Joule type losses estimate and the ways of their minimization**

Let us estimate the value of Joule type losses found in formula (22) and compare the calculated value with experimental data. We consider graphene nanoresonator with frequency $\omega_{res} \approx 130 MHz$ investigated in paper [17]. As for the case $m_{at} = 12 \cdot 1{,}67 \cdot 10^{-24}$, we have $m_{at} \cdot \omega^3 \approx 42 \, [g/s^3]$.

From formula (13) we get

$$E^0(\omega) = 3/4 \cdot \epsilon_1 / e \cdot \omega / V_F \approx$$

$$\approx 3/4 \cdot 3 \cdot 1{,}3/3 \, volt/cm = 3{,}9/4 \cdot 1/300 \, CGSE \tag{23}$$

The conductivity for our case was not written in [17] for graphene sample mentioned above. So we take it from another paper ([29]) where the parameters of experiment are close to the ones in [9]. From paper [29] for concentration value $n = 2{,}5 \cdot 10^{11} [cm^{-2}]$ we find in Fig.1 that $\sigma \approx 1{,}2 \cdot 10^9 \, [cm/s]$ (for good quality of the sample).

Estimate now the factor $a^2(I_x^2 + I_y^2)$ in (22). In [17] it is demonstrated that membrane oscillation critical amplitude after which nonlinearity appears is equal to $1{,}5 nm$. We assume it to be $h_{00} \approx 1 nm$. It is naturally to think that $\Delta a/a \approx h_{00}/h \approx 0{,}1$ i.e.

$$\eta_1/\eta_2 \approx \Delta a/h_{00} \approx (\Delta a/a) \cdot (a/h_{00}) \approx 2{,}5 \cdot 10^{-2}$$

Estimate the first term in the expression for $a^2(I_x^2 + I_y^2)$, using formulae (15), (16). Taking into consideration that graphene membrane surface has corrugations and assuming for simplicity the deformation height (depth) and the basis (length, width) to have close sizes we find

$$a^2 \cdot I_x^2 \approx (6{,}25 \cdot 10^{-4} a^4 \cdot (h_{xx}^2)^2 + \cdots) \approx (6{,}25/81) \cdot 10^{-8}$$

When estimating we assumed deformation radius to be $\delta_x \approx 15 nm$, and $\delta h/\delta_x \approx 2$. Other terms in formula for $a^2(I_x^2 + I_y^2)$ can be estimated similarly. Therefore we obtain

$$a^2 (I_x^2 + I_y^2) \approx 0{,}7 \cdot 10^{-8}. \tag{24}$$

Hence and from (22), (23) we find the approximate theoretical numerical value for Joule type losses in the sample mentioned above

$$Q_J^{-1} = \Delta E_J / E_{total} \approx 3 \cdot 10^{-5} \tag{25}$$

As experiment in [17] gives the result $Q \approx 14000$ we see that the Joule type losses are responsible for about 40 per cent and our model gives the reasonable magnitude of damping rate.

It is interesting that in paper [18] for the sample with about the same resonance frequency they obtained the quality factor $Q \approx 100\,000$. The measured increase of the quality factor to our point of view was obtained by authors as they used tension. From our formula (22) it is well seen that in this case the factor $(I_x^2 + I_y^2)$ is decreasing which enhances the quality factor, i.e. the measured increase of quality factor follows from our theory.

Now consider the question how one can minimize the Joule losses $Q_J^{-1}$. It is clear that the expressions $I_x, I_y$ in (15), (16), and consequently the losses (19) can be reduced by varying the form of the function $h(x, y)$ with the help of strains of different kinds. The fact that one can increase the quality factor by such actions was opened experimentally and it has become a subject of a new special field of activity which was called strain engineering. From the formula (22) the reason of this phenomenon is obvious.

One can decrease Joule losses also by switching on magnetic field perpendicular to graphene membrane plane. Indeed according to [30], in this case

$$\sigma_{xx} \approx \sigma(0)/[1 + (\Omega\tau)^2], \tag{26}$$

where

$$\sigma(0) = 2e^2 h^{-1} V_F \tau \sqrt{\pi n}, \tag{27}$$

and for the cyclotron frequency $\Omega$ we have

$$\Omega = V_F h^{-1} (\pi n)^{-1/2} eH/c. \qquad (28)$$

Here $n$ - electron concentration, $V_F$ - Fermi velocity, $\tau$ - relaxation time.

When $\Omega\tau \gg 1$ the value of $\sigma_{xx}(H)$ strongly decreases.

In paper [30] it is shown that one can decrease $\sigma$ by one order of magnitude with the help of magnetic field about 6 Tesla (see. Fig 1 in [30] and also [31]). This gives us possibility to decrease the damping (19) by one order of magnitude.

Note that the formula (26) was obtained using Boltzman equation and stops to be correct when quantization in magnetic field of Landau type starts. Nevertheless though the form of dependence changes the tendency of decreasing conserves.

Since the graphene membrane surface has corrugations the external magnetic field components parallel to vibrating membrane can arise. These components play the role of magnetomotive force. Hence as it is shown in [32-33] we can get extra damping. But as these components are very small compared to the perpendicular one we need not take them into account.

## *Conclusion*

In this paper we considered new dissipation mechanism for graphene nanoresonators i.e. Joule type losses caused by synthetic electric fields.

For the linear case (i.e. electromotive alternating force is rather small) the formulae for Joule type losses are obtained.

We would like to stress especially that in contrast to major part of papers dedicated to nanoresonators in which phenomenological approach within framework of continuum nonlinear elastic model (see [34] and last review-like paper [27]) was used (nonlinear Duffing oscillator) our results for Joule type losses are obtained on the basis of microscopic theory taking into account the specific features of graphene. Though the membrane vibration is supposed to be classical but the mechanism of losses in graphene nanoresonator is described within the framework of quantum solid state physics.

Using the obtained for Joule type losses formulae we calculated approximately their value. This estimate shows that their contribution to the general dissipation proved to be about 40 percents.

The possible methods of lowering down of Joule losses are as follows
- Application of strain engineering methods to minimize quantities $I_x$, $I_y$
- Switching on the magnetic field perpendicular to the graphene membrane.

Note that synthetic currents we considered in this paper lead not only to Joule type losses but cause dissipation due to their interaction with currents arising on gate. We hope to analyse these losses in the next paper.

In the present paper we found that the taking into account the inevitably existing in graphene membrane various corrugations gives essential contribution into the magnitude of the quality factor in graphene nanoresonator in megahertz and gigahertz frequency range. Obviously this mechanism should influence also a nonlinear electromagnetic response of graphene in terahertz and optical frequency range. In transport phenomena we should also take it into consideration. So for exact estimates by constructing different kinds of devices where we use graphene in nonstationary regime in such spheres as photonics optoelectronics etc. we should take into account the appearing of synthetic electric fields and their influence. Some ideas to minimize its negative action were also outlined.